\documentclass[11pt]{article}
\usepackage{amsmath,amssymb,amsfonts}
\usepackage[margin=2cm]{geometry}
\usepackage{microtype}
\usepackage{booktabs}
\usepackage[final]{graphicx} 
\usepackage{color}
\usepackage{tabularx}
\usepackage{array}
\usepackage{placeins}
\usepackage{rotating}
\usepackage{epstopdf}
\usepackage{pdflscape}
\usepackage{setspace}
\usepackage{enumerate}
\usepackage{float}
\usepackage{longtable}
\usepackage{caption}
\usepackage{textcomp}
\usepackage[utf8]{inputenc}
\usepackage{algorithm}
\usepackage{algpseudocode}
\usepackage[colorlinks=true, linkcolor=blue, citecolor=blue]{hyperref}
\usepackage{natbib}
\usepackage{hyperref}
\usepackage[symbol]{footmisc}

\usepackage[T1]{fontenc}
\usepackage{lmodern}

\title{A new composite Mann-Whitney test for two-sample survival comparisons with right-censored data}
\author{Abid Hussain$^{1,}\footnote{Corresponding author.}$ \ and Touqeer Ahmad$^2$  
\\
$^1$ Department of Statistics, Higher Education Department, Punjab, Pakistan \\ \href{mailto:abid0100@gmail.com}{abid0100@gmail.com} \\
$^2$ Institut Denis Poisson, Universit\'e d’Orl\'eans, France \\
\href{mailto:Email}{touqeer.ahmad@univ-orleans.fr} 
}

\date{}

\begin{document}
\maketitle
\begin{center}
\textbf{Abstract} 
\end{center}
A fundamental challenge in comparing two survival distributions with right-censored data is the selection of an appropriate nonparametric test, as the power of standard tests like the Log-rank and Wilcoxon is highly dependent on the often-unknown nature of the alternative hypothesis. This paper introduces a new, distribution-free two-sample test designed to overcome this limitation. The proposed method is based on a strategic decomposition of the data into uncensored and censored subsets, from which a composite test statistic is constructed as the sum of two independent Mann-Whitney statistics. This design allows the test to automatically and inherently adapt to various patterns of difference—including early, late, and crossing hazards—without requiring pre-specified parameters, pre-testing, or complex weighting schemes. An extensive Monte Carlo simulation study demonstrates that the proposed test robustly maintains the nominal Type I error rate. Crucially, its power is highly competitive with the optimal traditional tests in standard scenarios and superior in complex settings with crossing survival curves, while also exhibiting remarkable robustness to high levels of censoring. The test's power effectively approximates the maximum power achievable by either the Log-rank or Wilcoxon tests across a wide range of alternatives, offering a powerful, versatile, and computationally simple tool for survival analysis.	

\section{Introduction}
The two-sample comparison of survival curves is a cornerstone of statistical analysis in clinical trials, epidemiological studies, and reliability engineering. A pervasive challenge in this domain is the presence of right-censored data, where the exact time-to-event for some subjects remains unknown due to study termination or loss to follow-up \citep{Kleinbaum2012}. To evaluate the null hypothesis that the survival functions are identical, $H_{0}:S_{1}(t)=S_{2}(t)$, against a general alternative, a family of nonparametric tests employing weighting schemes is available in the statistical literature (see, for example, \citealt{Kalbfleisch2002}).

Among these, the Log-rank tests \citep{Mantel1966, Peto1972} and generalized Wilcoxon tests \citep{Gehan1965, Peto1972} are most prominent. Their power, however, is critically dependent on the unknown nature of the alternative hypothesis. The Log-rank procedure, which applies uniform weights across event times, is asymptotically optimal under the proportional hazards assumption, where survival differences manifest predominantly later in time. In contrast, Wilcoxon-type tests, which assign greater weight to early events, are generally more powerful for detecting early differences in survival (see, for example, \citealt{Harrington1982}). The challenge of test selection constitutes a critical dilemma for practitioners, wherein an inappropriate choice can compromise statistical power and lead to a Type II error. This issue has gained particular salience with the rising prevalence of non-proportional hazards (non-PH) in clinical data.

Non-PH patterns, particularly crossing survival curves where treatment effects change or reverse over time, are frequently encountered in modern clinical trials involving immunotherapies and targeted treatments \citep{Hess1994}. In such cases, both the Log-rank and Wilcoxon tests can suffer from a severe loss of power, potentially compromising the validity of a study's conclusions (see, for example, \citealt{Freidlin2002}). This limitation has motivated the development of more robust methods. One strategy involves flexible tests, such as the Fleming-Harrington $G^{\rho,\gamma}$ family \citep{Fleming1991}, which allow for targeting early, late, or middle differences through parameter selection. 
Another approach employs adaptive, data-driven strategies that use a pre-test or a weighted combination to choose a powerful test against various alternatives (see \cite{Lee1996} for a combination approach and \cite{Harrington1982} for a framework that generalizes and adapts to different types of alternatives). A significant practical limitation of these methods is that they often require the subjective selection of tuning parameters or introduce additional layers of complexity. Furthermore, they can inflate the Type I error rate if not meticulously calibrated, as the data-driven selection process invalidates standard asymptotic theory; for a comprehensive discussion of these issues, see the simulation study by \cite{Latta1981}.

Motivated by the need for a robust, powerful, and computationally simple test that requires no prior specification, we introduce a novel two-sample test for right-censored data. Our approach is philosophically aligned with the omnibus principle but is distinct in its construction. The proposed test is based on a novel decomposition of the data into uncensored and censored subsets. A composite test statistic is then formed from the sum of two independent Mann-Whitney statistics computed on these subsets. This structure inherently and automatically combines sensitivity to differences in both event times and censoring patterns, effectively adapting to a wide spectrum of alternatives without requiring pre-specified weights or a complex pre-testing procedure.

We demonstrate through an extensive Monte Carlo simulation study that the proposed test maintains the nominal Type I error rate and delivers a power profile that is highly competitive with the best traditional test in simple scenarios and superior in complex settings, such as those with crossing hazards. Notably, it exhibits remarkable robustness to high levels of censoring. Its power approximates the maximal power achievable by either the Log-rank or the Wilcoxon test across diverse alternatives, fulfilling the need for a stable, distribution-free, and versatile tool for analyzing survival data.

The remainder of this paper is structured as follows. Section~\ref{two-sample-s-test} provides a review of established nonparametric tests for comparing two survival distributions, including Gehan's generalized Wilcoxon test, the Cox-Mantel test, the Log-rank test, and the Peto-Peto test, thereby setting the stage for the proposed method. Section~\ref{proposed-test} introduces our novel test, detailing its motivation, the conceptual framework based on data decomposition, the formal construction of the composite statistic, and a discussion of its asymptotic properties and theoretical advantages. Section~\ref{sim-study} presents an extensive Monte Carlo simulation study, evaluating the test's control of Type I error and its power across a diverse range of alternative hypotheses, including challenging scenarios with crossing survival curves and high censoring rates. Section~\ref{real-app} demonstrates the practical utility of the proposed test through its application to four real-world datasets from clinical studies, comparing its performance with that of traditional tests. Finally, Section~\ref{concl} concludes the paper by summarizing the key findings and discussing the implications of our research.

\section{Two-sample survival tests}\label{two-sample-s-test}
Comparing survival experiences between two independent groups is a fundamental problem in survival analysis, particularly when the underlying distribution is unknown or the data are subject to right-censoring. Let $S_{1}(t)$ and $S_{2}(t)$ denote the survival functions for Group 1 and Group 2, respectively. The central hypothesis to be tested is the equality of these survival distributions, $H_{0}:S_{1}(t)=S_{2}(t)$ for all $t$, against the general alternative $H_{1}:S_{1}(t)\neq S_{2}(t)$. 

Consider two independent samples with sizes $n_1$ and $n_2$, yielding a total sample size of $N = n_1 + n_2$. Let $\{(Y_{1i},\delta_{1i})\}_{i=1}^{n_{1}}$ and $\{(Y_{2j},\delta_{2j})\}_{j=1}^{n_{2}}$ be the observed data for Group 1 and Group 2, respectively, where $Y_{ki}=\min(T_{ki}, C_{ki})$ is the observed time (the minimum of the true event time $T_{ki}$ and the censoring time $C_{ki}$) and $\delta_{ki}$ is the event indicator (1 for event, 0 for censored) for groups $k=1,2$.

\subsection{Gehan's generalized Wilcoxon test}
Gehan's test \citep{Gehan1965} extends the Mann-Whitney-Wilcoxon test \citep{Wilcoxon1945, Mann1947} to accommodate right-censoring and is particularly sensitive to early differences between survival curves. The test statistic is constructed from pairwise comparisons between observations from the two groups. For each pair $(Y_{1i},\delta_{1i})$ from Group 1 and $(Y_{2j},\delta_{2j})$ from Group 2, a score $U_{ij}$ is assigned as follows
\begin{equation}
U_{ij}=\begin{cases}
+1 & \text{if } Y_{1i} > Y_{2j} \text{ and } \delta_{2j} = 1 \\
+1 & \text{if } Y_{1i} = Y_{2j} \text{ and } \delta_{1i} = 1, \delta_{2j} = 0 \\
-1 & \text{if } Y_{1i} < Y_{2j} \text{ and } \delta_{1i} = 1 \\
-1 & \text{if } Y_{1i} = Y_{2j} \text{ and } \delta_{1i} = 0, \delta_{2j} = 1 \\
0 & \text{otherwise.}
\end{cases}
\end{equation}
The test statistic, $G_W$, is the sum over all possible $n_1 \times n_2$ pairs
\[G_W=\sum_{i=1}^{n_{1}}\sum_{j=1}^{n_{2}}U_{ij}.\]
Under the null hypothesis, $E(G_W)=0$. The variance of $G_W$, accounting for ties and censoring, is given by
\[\text{Var}(G_W)=\frac{n_{1}n_{2}}{N(N-1)}\sum_{i=1}^{N}(R_{1i}-R_{2i})^{2},\]
where \(R_{1i}\) equals one plus the number of observations for which the focal observation is the larger value, effectively its rank from the minimum, and \(R_{2i}\) equals one plus the number of observations that are larger than the focal observation, which relates to its rank from the maximum (see, for example, \citealt{Lee2003}). The standardized test statistic
\[Z=\frac{G_W}{\sqrt{\text{Var}(G_W)}},\]
is asymptotically distributed as a standard normal variate under $H_{0}$.

\subsection{The Cox-Mantel test}
The Cox-Mantel test \citep{Mantel1966} is a fundamental nonparametric procedure formulated based on the hypergeometric distribution at each distinct event time. \noindent Consider two treatment groups with combined distinct ordered failure times
\[t_{(1)} < t_{(2)} < \cdots < t_{(m)}.\]
Let $m_{(i)}$ denote the number of failures occurring at time $t_{(i)}$, satisfying
\[\sum_{i=1}^{k} m_{(i)} = f_1 + f_2,\]
where $f_1$ and $f_2$ represent total failures in groups 1 and 2, respectively.
\medskip\noindent
Define the risk set $R(t)$ as all subjects surviving and uncensored just prior to time $t$. Let $f_{1t}$ and $f_{2t}$ denote subjects in $R(t)$ from groups 1 and 2, with total risk set size at $t_{(i)}$ being \(r_{(i)} = f_{1t} + f_{2t}\).
\medskip\noindent
Let $P_{(i)} = f_{2t}/r_{(i)}$ represent the proportion of the risk set from group 2 at $t_{(i)}$. The test statistic is defined as
\[C = f_2 - \sum_{i=1}^m m_{(i)} P_{(i)}. \]
Under null hypothesis, \(E(C)=0\), and the variance is given by
\[Var(C) = \sum_{i=1}^m \frac{m_{(i)}(r_{(i)} - m_{(i)})}{r_{(i)} - 1} \cdot P_{(i)}(1 - P_{(i)}).\]
\medskip\noindent
For further elaboration, we refer to \cite{Lee2003}. Under the identical survival distributions, the standardized statistic, \(Z = \frac{C}{\sqrt{Var(C}}\), follows approximately a standard normal distribution. This provides the basis for the asymptotic two-sample test (see, for example, \citealt{Cox1972}).

\subsection{The Log-rank test}
The Log-rank test uses scores based on the logarithm of the survival function, building upon \cite{Mantel1966} generalization of the earlier \cite{Savage1956} test. An alternative formulation of the Log-rank test uses a chi-square framework to assess the discrepancy between observed and expected failure counts under the null hypothesis of identical survival. Denote by $\mathcal{O}_1$ and $\mathcal{O}_2$ the total observed failures in groups 1 and 2, and by $\mathcal{E}_1$ and $\mathcal{E}_2$ the corresponding expected failures. The test statistic
\[X^2 = \frac{(\mathcal{O}_1 - \mathcal{E}_1)^2}{\mathcal{E}_1} + \frac{(\mathcal{O}_2 - \mathcal{E}_2)^2}{\mathcal{E}_2},\]
approximately follows a chi-square distribution with 1 degree of freedom. Significant evidence against the null hypothesis occurs when $X^2$ exceeds the critical value (e.g., $\geq \chi_{1,0.05}^2$ for $\alpha = 0.05$), indicating differential treatment effectiveness.

The calculation of expected counts proceeds by considering each distinct failure time. Let $d_t$ represent the number of failures at time $t$, with $n_{1t}$ and $n_{2t}$ denoting the numbers of subjects at risk in each group just prior to time $t$. The expected failures for each group at time $t$ are
\[
e_{1t} = \frac{n_{1t}}{n_{1t} + n_{2t}} \times d_t \qquad e_{2t} = \frac{n_{2t}}{n_{1t} + n_{2t}} \times d_t.\]
Cumulative expected counts are obtained by summing over all failure times
\[\mathcal{E}_1 = \sum_{t} e_{1t} \qquad \mathcal{E}_2 = \sum_{t} e_{2t}.\]

\subsection{Peto and Peto's generalized Wilcoxon test}
\citet{Peto1972} proposed a generalization of the Wilcoxon test by using the \cite{Kaplan1958} survival estimate, $\hat{S}(t)$, of the pooled sample for explicit weighting. For each observation
\begin{itemize}
\item For an uncensored observation at time $t$: $\mu_{i}=\hat{S}(t-)+\hat{S}(t+)-1$,
\item For a censored observation at time $T$: $\mu_{i}=\hat{S}(T)-1$.
\end{itemize}
The test statistic is the sum of scores for Group 1
\[W_{PP}=\sum_{j=1}^{n_{1}}\mu_{1j}.\]
Under $H_{0}$, $E(W_{PP})=0$, and its variance is
\[
\text{Var}(W_{PP})=\frac{n_{1}n_{2}}{N(N-1)}\sum_{i=1}^{N}\mu_{i}^{2},
\]
where $\mu_{i}$ are the scores for all subjects in the pooled sample (see for further elaboration, \citealt{Lee2003}). The standardized statistic
\[Z=\frac{W_{PP}}{\sqrt{\text{Var}(W_{PP})}},\]
follows an asymptotic standard normal distribution under $H_{0}$.

\section{Proposed test}\label{proposed-test}
\subsection{Motivation and Conceptual Framework}
The performance of existing tests is intrinsically linked to the nature of the alternative hypothesis. The Log-rank test is optimal for proportional hazards (late differences), while Mann-Whitney-Wilcoxon-type tests are more powerful for early differences. This creates a significant dilemma for practitioners, as the true nature of the difference is unknown a priori. In complex scenarios, particularly when survival curves intersect, both classes of tests can suffer a severe loss of power. While adaptive or pre-test strategies have been proposed to select the best test post-hoc, they introduce additional complexity, and their performance can be sensitive to the pre-test itself. Therefore, a unified, single-statistic test that automatically adapts to various alternative patterns without requiring a pre-test is highly desirable.

The core insight of our proposed test is that the informational content regarding the difference between two survival distributions is encoded differently in the uncensored observations, which provide precise event times, and the censored observations, which provide a lower bound for the event time. By strategically decomposing the data and constructing a statistic that leverages both sources of information independently, we can create a test that is responsive to a wider range of alternatives.

\subsection{Test formulation and asymptotic properties}
Let the complete two-sample right-censored data be denoted by $\mathcal{D}=\{(Y_{ij},\delta_{ij}):i=1,2;j=1,\ldots,n_{i}\}$, where $Y_{ij}=\min(T_{ij},C_{ij})$ is the observed time and $\delta_{ij}$ is the event indicator. We propose a partition of $\mathcal{D}$ into two statistically independent subsets: the uncensored subsample, $\mathcal{D}_{U}=\{(Y_{ij},\delta_{ij})\in\mathcal{D} \mid \delta_{ij}=1\}$, with sizes $n_{1U}$ and $n_{2U}$ (total $N_{U}$), and the censored subsample, $\mathcal{D}_{C}=\{(Y_{ij},\delta_{ij})\in\mathcal{D} \mid \delta_{ij}=0\}$, with sizes $n_{1C}$ and $n_{2C}$ (total $N_{C}$).

Our test statistic, $U_{P}$, is constructed as a composite Mann-Whitney statistic
\begin{equation}
U_{P}=U_{U}+U_{C}=\sum_{i\in\mathcal{D}_{1U}}\sum_{j\in\mathcal{D}_{2U}}I(Y_{1i}>Y_{2j})+\sum_{i\in\mathcal{D}_{1C}}\sum_{j\in\mathcal{D}_{2C}}I(Y_{1i}>Y_{2j}),
\end{equation}
where \(\mathcal{D}_{iU}\) and \(\mathcal{D}_{iC}\) constitute a partition of the data for group \(i\), representing the respective subsets of uncensored and censored observations. This formulation offers a distinct advantage. The component $U_{U}$, from the uncensored subsample, is powerful against alternatives where the actual event times differ, making it sensitive to a wide range of patterns, including crossing hazards. The component, $U_{C}$, from the censored subsample incorporates information from the censoring patterns, where a systematic difference can itself be indicative of an underlying difference in survival, especially with informative censoring or when differences manifest in the risk sets over time. By summing these independent components, $U_{P}$ aggregates evidence of stochastic ordering from both the event data and the censoring process.

Under the null hypothesis $H_{0}:S_{1}(t)=S_{2}(t)$ with non-informative censoring, all observations are independent and identically distributed. The expectation of the proposed statistic is simply the sum of the expectations of two independent Mann-Whitney statistics
\begin{equation*}
E\big(U_{P} \mid H_{0}\big) = E(U_{U}) + E(U_{C}) = \frac{n_{1U}n_{2U} + n_{1C}n_{2C}}{2}.
\end{equation*}
Given the independence of $U_{U}$ and $U_{C}$, the variance is the sum of their variances. For a Mann-Whitney statistic with group sizes $n_1$ and $n_2$, the variance under $H_0$ is $\frac{n_1n_2(n_1+n_2+1)}{12}$. Applying this
\begin{equation*}
\text{Var}(U_{P} \mid H_{0}) = \text{Var}(U_{U}) + \text{Var}(U_{C}) = \frac{n_{1U}n_{2U}(N_{U}+1) + n_{1C}n_{2C}(N_{C}+1)}{12}.
\end{equation*}
The standardized test statistic is then
\begin{equation}
Z=\frac{U_{P}-E[U_{P}]}{\sqrt{\text{Var}(U_{P})}},
\end{equation}
which, by the Central Limit Theorem, converges in distribution to a standard normal variate under $H_{0}$ as $\min(n_{1}, n_{2}) \to \infty$.

The proposed test possesses several key strengths. It is distribution-free, relying on no parametric assumptions. It automatically adapts to the data, as the relative contribution of $U_{U}$ and $U_{C}$ shifts naturally without requiring pre-specified parameters. Its structure enables it to maintain power in challenging scenarios, such as crossing hazards, as validated empirically in Section \ref{sim-study}. It remains computationally straightforward, relying on simple pairwise comparisons within well-defined subsets.

\section{Monte Carlo simulation study}\label{sim-study}
An extensive Monte Carlo simulation study is conducted to empirically evaluate the performance of the proposed test. The primary objectives are to verify that the test maintains the nominal Type I error rate under the null hypothesis and to compare its power with that of established nonparametric tests—namely, the Gehan, Cox-Mantel, Log-rank, and Peto-Peto tests—under various alternative scenarios. All simulations and analyses are performed using the \texttt{R} statistical environment (\texttt{Version 4.5.1}). For each experimental configuration, 10,000 independent datasets are generated, and the proportion of null hypothesis rejections at the \(\alpha=0.05\) significance level is recorded.

\subsection{Size of the tests}
The control of the Type I error rate is assessed by generating data under the null hypothesis of identical survival distributions for both groups. Data are simulated from four common survival distributions: Exponential(1), Weibull(1,1), Log-logistic(1,0.5), and Gamma(1,2). To introduce right-censoring, the approach of \cite{Letón2005} is adopted, which utilizes uniform distributions, \(U(0, \theta)\). The parameter \(\theta\) is adjusted to achieve censoring percentages ranging from 0\% to approximately 79\%. The censoring mechanism is identical for both groups in all configurations.

The empirical sizes for the scenario with sample sizes \(n_1 = n_2 = 50 \) are summarized in Table~\ref{tab1_cen}. For 10,000 replications, the 95\% acceptance region for the nominal \(\alpha=0.05\) level, formally defined by the interval
\begin{equation*}
\bigg(0.049-z_{0.975} \sqrt{\frac{0.05 \times 0.95}{10,000}},~ 0.051+z_{0.975} \sqrt{\frac{0.05 \times 0.95}{10,000}}\bigg)=\left(0.0447, 0.0553\right),
\end{equation*}
where \(z_{0.975}\) is the \textit{97.5}th percentile of the standard normal distribution. As shown in Table~\ref{tab1_cen}, all tests demonstrate excellent control of the Type I error rate, with all estimated sizes falling comfortably within this acceptable range across all distributions and censoring levels. For instance, the Gehan test's size ranged from 0.0457 to 0.0573, the Log-rank test from 0.0448 to 0.0546, and crucially, the proposed test also maintained a valid size, with estimates ranging from 0.0462 to 0.0528. This confirms that the asymptotic normal approximation for the proposed test statistic provides an accurate reference distribution under the null hypothesis, ensuring that any observed power differences are attributable to the tests relative efficiencies and not to size distortions.

\subsection{Power of the tests}
\begin{figure}[!t]
\centering
\resizebox*{18cm}{!}{\includegraphics{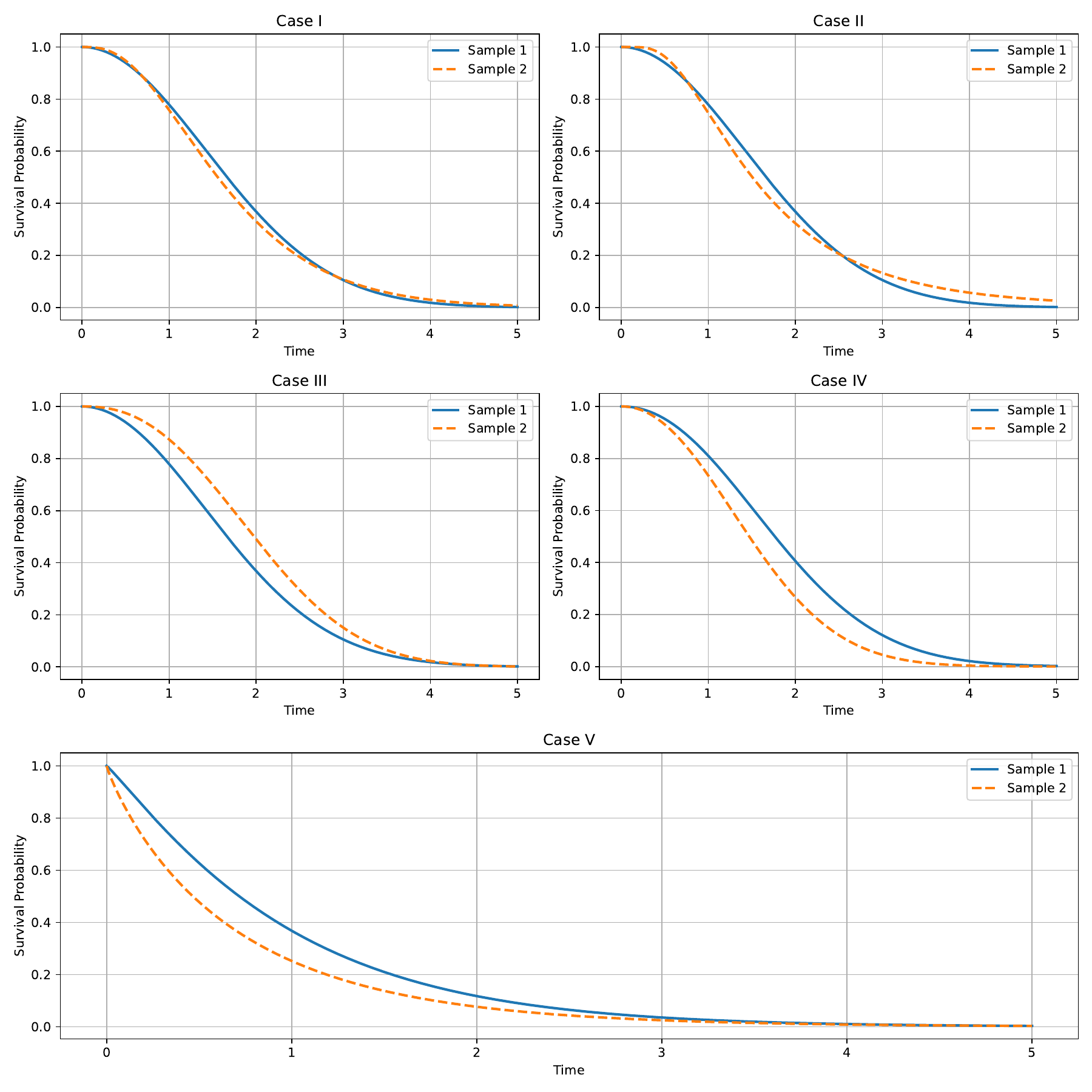}}
\caption{Survival distributions for the Cases I-V, defined in Table~\ref{tab3-cases-cen}.}
\label{fig:cases-obs}
\end{figure}

A simulation study is conducted to evaluate the power of the tests under various alternatives to the null hypothesis. The framework, adapted from \cite{Philonenko2015}, features five complex survival scenarios (Case-I to Case-V) generated from Weibull, Gamma, and Lognormal distributions (see Table~\ref{tab2-power}). This design is selected to represent a challenging and diverse spectrum of survival patterns. This can be observed in Figure~\ref{fig:cases-obs} as Case-I and II feature two crossing points, creating scenarios where early and late differences can cancel out; Case-III represents stochastically ordered survival functions with early differences; Case-IV represents stochastically ordered functions with late differences; and Case-V features a single, very late crossing point with dominant early differences. Censoring is introduced using independent Weibull distributions for each group, with parameters adjusted to achieve target censoring rates from 0\% to 50\%, as specified in Table~\ref{tab3-cases-cen}. The simulated power for all tests across these five cases, for sample sizes of (50, 50), (100, 100), and (200, 200), is presented in Tables~\ref{tab4-power}--\ref{tab8-power}.

\begin{table}[!t]
\centering
\setlength{\tabcolsep}{3pt} 
\caption{Size of the tests for \(n_1=n_2=50\) and censoring \(U(0,\theta)\).}
\begin{tabular}{@{}lccccccccccc@{}}
\hline
&&&&&&&Test&&&&\\ 
\cline{4-12}
Dist.&\(\theta\)&Censoring (\%)&Gehan&&Cox-Mantel&&Log-rank&&Peto-Peto&&Proposed\\ \hline
\(\text{Exp}(1)\)&--&0&0.0496&&0.0530&&0.0486&&0.0496&&0.0480\\
&4&25&0.0513&&0.0515&&0.0475&&0.0500&&0.0498\\
&2&43&0.0488&&0.0507&&0.0478&&0.0477&&0.0498\\
&1&63&0.0518&&0.0507&&0.0493&&0.0523&&0.0474\\
&0.5&79&0.0565&&0.0550&&0.0538&&0.0552&&0.0473\\
&&&&&&&&&&&\\
\(\text{Weibull}(1,1)\)&--&0&0.0532&&0.0582&&0.0513&&0.0532&&0.0525\\
&4&25&0.0525&&0.0557&&0.0520&&0.0518&&0.0485\\
&2&43&0.0522&&0.0518&&0.0492&&0.0515&&0.0497\\
&1&63&0.0573&&0.0562&&0.05461&&0.0555&&0.0528\\
&0.5&79&0.0493&&0.0490&&0.0478&&0.0482&&0.0503\\
&&&&&&&&&&&\\
\(\text{Log-Logistic}(1,0.5)\)&--&0&0.0493&&0.0496&&0.0448&&0.0493&&0.0482\\
&4&28&0.0505&&0.0553&&0.0530&&0.0530&&0.0492\\
&2&40&0.0541&&0.0527&&0.0505&&0.0543&&0.0493\\
&1&55&0.0515&&0.0533&&0.0513&&0.0508&&0.0471\\
&0.5&69&0.0480&&0.0491&&0.0477&&0.0493&&0.0525\\
&&&&&&&&&&&\\
\(\text{Gamma}(1,2)\)&--&0&0.0457&&0.0508&&0.0455&&0.0458&&0.0488\\
&4&12&0.0491&&0.0543&&0.0500&&0.0483&&0.0501\\
&2&25&0.0488&&0.0525&&0.0488&&0.0505&&0.0463\\
&1&43&0.0503&&0.0538&&0.0512&&0.0496&&0.0498\\
&0.5&63&0.0544&&0.0526&&0.0491&&0.0591&&0.0462\\
\hline
\end{tabular}
\label{tab1_cen}%
\end{table}

The results reveal a consistent and compelling performance profile for the proposed test. In the challenging scenarios with crossing survival curves (Case-I and II), where overall power is low, the proposed test is highly competitive. It often matched the performance of the Wilcoxon-based tests at lower censoring levels and, importantly, demonstrated a significant advantage at higher censoring levels. For example, in Case-II with \(n_1=n_2=200\) and 20\% censoring (Table~\ref{tab5-power}), the proposed test's power (0.2179) substantially exceeded that of the next best test, Gehan (0.1430).

\begin{table}[!t]
\centering
\caption{Distributions for power comparison with different configurations of intersections.}
\begin{tabular}{llll}
\hline
Case & Sample-1 & Sample-II & Points of intersections\\ \hline
I    & \(f_{We}(0, 2.0, 2.0)\) & \(f_\Gamma(0.557706, 3.12154)\) & 2: 0.67, 2.90 \\
II   & \(f_{We}(0, 2.0, 2.0)\) & \(f_{LgN}(0.4096, 0.6179)\) & 2: 0.75, 2.55\\
III  & \(f_{We}(0, 2.0, 2.0)\) & \(f_{We}(0, 2.3, 2.4)\)     & 0\\
IV   & \(f_{We}(0, 2.1, 2.1)\) & \(f_{We}(0, 1.75, 2.1)\) & 0 \\
V    & \(f_{We}(0, 1.0, 1.1)\) & \(f_{We}(0, 0.7, 0.9)\)    & 1: 4.98\\
\hline
\end{tabular}
\label{tab2-power}%
\end{table} 

\begin{table}[!t]
\centering
\caption{Distributions of censored times.}
\begin{tabular}{ccccc}
\hline
$\text{Case}$ & Censored rate (\%) & Sample-I: \(C_1 \sim f_{We}(\mu, \sigma, \lambda)\) & & Sample-II: \(C_2 \sim f_{We}(\mu, \sigma, \lambda)\) \\
\hline
I& 10& (0.000, 5.795, 2.000) & & (0.000, 5.847, 2.000) \\
& 20 & (0.000, 3.992, 2.000) & & (0.000, 3.853, 2.000) \\
& 30 & (0.000, 3.050, 2.000) & & (0.000, 2.985, 2.000) \\
& 40 & (0.000, 2.426, 2.000) & & (0.000, 2.385, 2.000) \\
& 50 & (0.000, 2.000, 2.000) & & (0.000, 1.909, 2.000) \\
&&&&\\
II&10& (0.000, 5.795, 2.000) & & (0.410, 5.220, 2.000) \\
& 20 & (0.000, 3.992, 2.000) & & (0.410, 3.253, 2.000) \\
& 30 & (0.000, 3.050, 2.000) & & (0.410, 2.373, 2.000) \\
& 40 & (0.000, 2.426, 2.000) & & (0.410, 1.776, 2.000) \\
& 50 & (0.000, 2.000, 2.000) & & (0.410, 1.371, 2.000) \\
&&&&\\
III&10&(0.000, 5.795, 2.000) & & (0.000, 3.736, 6.340) \\
& 20 & (0.000, 3.992, 2.000) & & (0.000, 4.535, 2.000) \\
& 30 & (0.000, 3.050, 2.000) & & (0.000, 3.538, 2.000) \\
& 40 & (0.000, 2.426, 2.000) & & (0.000, 2.805, 2.000) \\
& 50 & (0.000, 2.000, 2.000) & & (0.000, 2.369, 2.000) \\
&&&&\\
IV&10& (0.000, 6.227, 2.000) & & (0.000, 5.301, 2.000) \\
& 20 & (0.000, 4.240, 2.000) & & (0.000, 3.463, 2.000) \\
& 30 & (0.000, 3.227, 2.000) & & (0.000, 2.688, 2.000) \\
& 40 & (0.000, 2.593, 2.000) & & (0.000, 2.162, 2.000) \\
& 50 & (0.000, 2.090, 2.000) & & (0.000, 1.769, 2.000) \\
&&&&\\
V &10& (0.000, 3.724, 2.000) & & (0.000, 2.819, 2.000) \\
& 20 & (0.000, 2.249, 2.000) & & (0.000, 1.745, 2.000) \\
& 30 & (0.000, 1.629, 2.000) & & (0.000, 1.164, 2.000) \\
& 40 & (0.000, 1.196, 2.000) & & (0.000, 0.813, 2.000) \\
& 50 & (0.000, 0.883, 2.000) & & (0.000, 0.582, 2.000) \\
\hline
\end{tabular}
\label{tab3-cases-cen}%
\end{table}
When differences are prominent early in time (Case-III and V), the Gehan and Peto-Peto tests are, as expected, the most powerful procedures at 0\% censoring. However, a key strength of the proposed test emerged as the censoring rate increased. Its power not only remained robust but often increased, eventually surpassing all other tests. In Case-III with \(n_1=n_2=200\) and 50\% censoring (Table~\ref{tab6-power}), the proposed test achieved a power of 0.9393, significantly higher than Gehan (0.8311) and Log-rank (0.6920). This trend is even more pronounced in Case-V (Table~\ref{tab8-power}), where the proposed test's power reached 1.0000 under high censoring, showcasing exceptional stability and sensitivity.

In the scenario with late differences (Case-IV), which favors the Log-rank test, the proposed test performed commendably. While the Log-rank test is the most powerful procedure at low censoring, the proposed test demonstrated superior robustness to censoring compared to the Wilcoxon-type tests. As shown in Table~\ref{tab7-power}, it maintained substantially higher power as the censoring rate increased (e.g., 0.8666 vs. 0.6382 for Gehan at 50\% censoring and \(n_1=n_2=200\)). 

\begin{table}[!t]
\centering
\setlength{\tabcolsep}{3pt} 
\caption{Simulated power of competing tests for Case-I.}
\begin{tabular}{@{}cccccccccccccccc@{}}
\hline
&&&&&&&&&Test&&&&&&\\ 
\cline{4-16}
\(n_1\)&\(n_2\)&Censoring (\%)&Gehan&&&Cox-Mantel&&&Log-rank&&&Peto-Peto&&&Proposed\\ \hline
50&50&0&0.0712&&&0.0585&&&0.0518&&&0.0712&&&0.0700\\
&&10&0.0704&&&0.0606&&&0.0548&&&0.0677&&&0.0650\\
&&20&0.0660&&&0.0602&&&0.0536&&&0.0686&&&0.0721\\
&&30&0.0687&&&0.0642&&&0.0577&&&0.0664&&&0.0695\\
&&40&0.0658&&&0.0673&&&0.0615&&&0.0666&&&0.0633\\
&&50&0.0610&&&0.0593&&&0.0542&&&0.0612&&&0.0707\\
&&&&&&&&&&&&&&&\\
100&100&0&0.0797&&&0.0532&&&0.0502&&&0.0797&&&0.0789\\
&&10&0.0860&&&0.0600&&&0.0562&&&0.0857&&&0.0864\\
&&20&0.0733&&&0.0653&&&0.0608&&&0.0746&&&0.0820\\
&&30&0.0846&&&0.0718&&&0.0676&&&0.0849&&&0.0869\\
&&40&0.0750&&&0.0725&&&0.0675&&&0.0780&&&0.0655\\
&&50&0.0656&&&0.0696&&&0.0662&&&0.0691&&&0.0842\\
&&&&&&&&&&&&&&&\\
200&200&0&0.1125&&&0.0515&&&0.0498&&&0.1125&&&0.1115\\
&&10&0.1130&&&0.0678&&&0.0667&&&0.1132&&&0.1135\\
&&20&0.1075&&&0.0711&&&0.0690&&&0.1078&&&0.1297\\
&&30&0.1004&&&0.0788&&&0.0776&&&0.1033&&&0.1170\\
&&40&0.0971&&&0.0869&&&0.0845&&&0.0991&&&0.0931\\
&&50&0.0838&&&0.0947&&&0.0924&&&0.0942&&&0.1229\\
\hline
\end{tabular}
\label{tab4-power}%
\end{table} 

\begin{table}[!t]
\centering
\setlength{\tabcolsep}{3pt} 
\caption{Simulated power of competing tests for Case-II.}
\begin{tabular}{@{}cccccccccccccccc@{}}
\hline
&&&&&&&&&Test&&&&&&\\ 
\cline{4-16}
\(n_1\)&\(n_2\)&Censoring (\%)&Gehan&&&Cox-Mantel&&&Log-rank&&&Peto-Peto&&&Proposed\\ \hline
50&50&0&0.0693&&&0.0485&&&0.0415&&&0.0693&&&0.0689\\
&&10&0.0772&&&0.0535&&&0.0488&&&0.0750&&&0.0813\\
&&20&0.0720&&&0.0572&&&0.0527&&&0.0710&&&0.0887\\
&&30&0.0771&&&0.0663&&&0.0612&&&0.0745&&&0.0788\\
&&40&0.0725&&&0.0643&&&0.0602&&&0.0727&&&0.0638\\
&&50&0.0642&&&0.0743&&&0.0642&&&0.0683&&&0.0570\\
&&&&&&&&&&&&&&&\\
100&100&0&0.0860&&&0.0567&&&0.0515&&&0.0860&&&0.0845\\
&&10&0.0893&&&0.0470&&&0.0433&&&0.0867&&&0.1103\\
&&20&0.0985&&&0.0493&&&0.0480&&&0.0940&&&0.1351\\
&&30&0.1052&&&0.0712&&&0.0671&&&0.1012&&&0.1079\\
&&40&0.0938&&&0.0840&&&0.0790&&&0.0975&&&0.0822\\
&&50&0.0843&&&0.0877&&&0.0815&&&0.0903&&&0.0808\\
&&&&&&&&&&&&&&&\\
200&200&0&0.1257&&&0.0737&&&0.0672&&&0.1257&&&0.1248\\
&&10&0.1408&&&0.0455&&&0.0435&&&0.1342&&&0.1778\\
&&20&0.1430&&&0.0547&&&0.0520&&&0.1318&&&0.2179\\
&&30&0.1470&&&0.0755&&&0.0741&&&0.1375&&&0.1665\\
&&40&0.1347&&&0.1020&&&0.0990&&&0.1342&&&0.1515\\
&&50&0.1260&&&0.1387&&&0.1328&&&0.1460&&&0.1547\\
\hline
\end{tabular}
\label{tab5-power}%
\end{table} 

\begin{table}[ht]
\centering
\setlength{\tabcolsep}{3pt} 
\caption{Simulated power of competing tests for Case-III.}
\begin{tabular}{@{}cccccccccccccccc@{}}
\hline
&&&&&&&&&Test&&&&&&\\ 
\cline{4-16}
\(n_1\)&\(n_2\)&Censoring (\%)&Gehan&&&Cox-Mantel&&&Log-rank&&&Peto-Peto&&&Proposed\\ \hline
50&50&0&0.3290&&&0.2172&&&0.2062&&&0.3290&&&0.3263\\
&&10&0.3261&&&0.2392&&&0.2293&&&0.3237&&&0.2885\\
&&20&0.3343&&&0.2453&&&0.2335&&&0.3263&&&0.3652\\
&&30&0.3235&&&0.2473&&&0.2315&&&0.3140&&&0.3965\\
&&40&0.3193&&&0.2533&&&0.2395&&&0.3037&&&0.3958\\
&&50&0.3131&&&0.2562&&&0.2380&&&0.2975&&&0.4197\\
&&&&&&&&&&&&&&&\\
100&100&0&0.5785&&&0.3817&&&0.3732&&&0.5785&&&0.5768\\
&&10&0.5923&&&0.4148&&&0.4065&&&0.5876&&&0.5235\\
&&20&0.5636&&&0.3990&&&0.3913&&&0.5502&&&0.6263\\
&&30&0.5611&&&0.4237&&&0.4130&&&0.5458&&&0.6798\\
&&40&0.5513&&&0.4188&&&0.4070&&&0.5247&&&0.6602\\
&&50&0.5327&&&0.4265&&&0.4153&&&0.5127&&&0.6838\\
&&&&&&&&&&&&&&&\\
200&200&0&0.8708&&&0.6421&&&0.6390&&&0.8708&&&0.8703\\
&&10&0.8668&&&0.6668&&&0.6618&&&0.8618&&&0.8082\\
&&20&0.8593&&&0.6823&&&0.6782&&&0.8477&&&0.9108\\
&&30&0.8600&&&0.6888&&&0.6830&&&0.8393&&&0.9337\\
&&40&0.8412&&&0.6973&&&0.6906&&&0.8248&&&0.9302\\
&&50&0.8311&&&0.7003&&&0.6920&&&0.8097&&&0.9393\\
\hline
\end{tabular}
\label{tab6-power}%
\end{table}

\begin{table}[!t]
\centering
\setlength{\tabcolsep}{3pt} 
\caption{Simulated power of competing tests for Case-IV.}
\begin{tabular}{@{}cccccccccccccccc@{}}
\hline
&&&&&&&&&Test&&&&&&\\ 
\cline{4-16}
\(n_1\)&\(n_2\)&Censoring (\%)&Gehan&&&Cox-Mantel&&&Log-rank&&&Peto-Peto&&&Proposed\\ \hline
50&50&0&0.3747&&&0.4590&&&0.4430&&&0.3747&&&0.3725\\
&&10&0.3380&&&0.4190&&&0.3993&&&0.3462&&&0.3383\\
&&20&0.3093&&&0.3885&&&0.3655&&&0.3275&&&0.3415\\
&&30&0.2813&&&0.3495&&&0.3265&&&0.3037&&&0.3390\\
&&40&0.2498&&&0.3148&&&0.2921&&&0.2808&&&0.3433\\
&&50&0.2068&&&0.2631&&&0.2461&&&0.2403&&&0.3258\\
&&&&&&&&&&&&&&&\\
100&100&0&0.6410&&&0.7657&&&0.7575&&&0.6410&&&0.6393\\
&&10&0.5977&&&0.7096&&&0.7000&&&0.6132&&&0.5988\\
&&20&0.5495&&&0.6688&&&0.6558&&&0.5782&&&0.6039\\
&&30&0.5063&&&0.6107&&&0.5957&&&0.5448&&&0.5978\\
&&40&0.4303&&&0.5332&&&0.5158&&&0.4747&&&0.6015\\
&&50&0.3815&&&0.4835&&&0.4683&&&0.4406&&&0.5867\\
&&&&&&&&&&&&&&&\\
200&200&0&0.9107&&&0.9642&&&0.9638&&&0.9107&&&0.9102\\
&&10&0.8788&&&0.9500&&&0.9475&&&0.8910&&&0.8738\\
&&20&0.8458&&&0.9233&&&0.9192&&&0.8662&&&0.8827\\
&&30&0.7955&&&0.8822&&&0.8762&&&0.8248&&&0.8778\\
&&40&0.7143&&&0.8277&&&0.8215&&&0.7685&&&0.8869\\
&&50&0.6382&&&0.7598&&&0.7526&&&0.7139&&&0.8666\\
\hline
\end{tabular}
\label{tab7-power}%
\end{table}

\begin{table}[!t]
\centering
\setlength{\tabcolsep}{3pt} 
\caption{Simulated power of competing tests for Case-V.}
\begin{tabular}{@{}cccccccccccccccc@{}}
\hline
&&&&&&&&&Test&&&&&&\\ 
\cline{4-16}
\(n_1\)&\(n_2\)&Censoring (\%)&Gehan&&&Cox-Mantel&&&Log-rank&&&Peto-Peto&&&Proposed\\ \hline
50&50&0&0.4107&&&0.2822&&&0.2710&&&0.4107&&&0.4073\\
&&10&0.4250&&&0.3123&&&0.3016&&&0.4234&&&0.5067\\
&&20&0.4215&&&0.3281&&&0.3166&&&0.4174&&&0.5299\\
&&30&0.4110&&&0.3300&&&0.3128&&&0.4042&&&0.6408\\
&&40&0.4052&&&0.3573&&&0.3396&&&0.4029&&&0.7630\\
&&50&0.4135&&&0.3610&&&0.3397&&&0.4010&&&0.8433\\
&&&&&&&&&&&&&&&\\
100&100&0&0.7020&&&0.5031&&&0.4945&&&0.7020&&&0.7008\\
&&10&0.7032&&&0.5320&&&0.5233&&&0.6988&&&0.7983\\
&&20&0.7036&&&0.5621&&&0.5552&&&0.6959&&&0.8273\\
&&30&0.7019&&&0.5896&&&0.5778&&&0.6927&&&0.9103\\
&&40&0.6961&&&0.6003&&&0.5846&&&0.6849&&&0.9687\\
&&50&0.6776&&&0.6110&&&0.5955&&&0.6656&&&0.9893\\
&&&&&&&&&&&&&&&\\
200&200&0&0.9403&&&0.7622&&&0.7583&&&0.9403&&&0.9402\\
&&10&0.9413&&&0.8067&&&0.8030&&&0.9386&&&0.9739\\
&&20&0.9432&&&0.8497&&&0.8471&&&0.9393&&&0.9843\\
&&30&0.9401&&&0.8680&&&0.8627&&&0.9362&&&0.9970\\
&&40&0.9325&&&0.8741&&&0.8674&&&0.9250&&&0.9997\\
&&50&0.9298&&&0.8830&&&0.8725&&&0.9227&&&1.0000
\\\hline
\end{tabular}
\label{tab8-power}%
\end{table}

In summary, the simulation study demonstrates that the proposed test successfully controls the Type I error rate and exhibits a highly desirable power profile. It automatically adapts to the nature of the alternative hypothesis, providing power that is competitive with the best-performing traditional test in simple scenarios, superior in complex scenarios with crossing survival curves, and remarkably robust to high levels of right-censoring. This performance aligns with the test's design philosophy of aggregating evidence from both uncensored and censored observations to achieve a stable and powerful omnibus property.

\section{Real-life data examples}\label{real-app}
\begin{figure}[!t]
\centering
\begin{tabular}{c c}
\includegraphics[width=8cm, height=6.5cm]{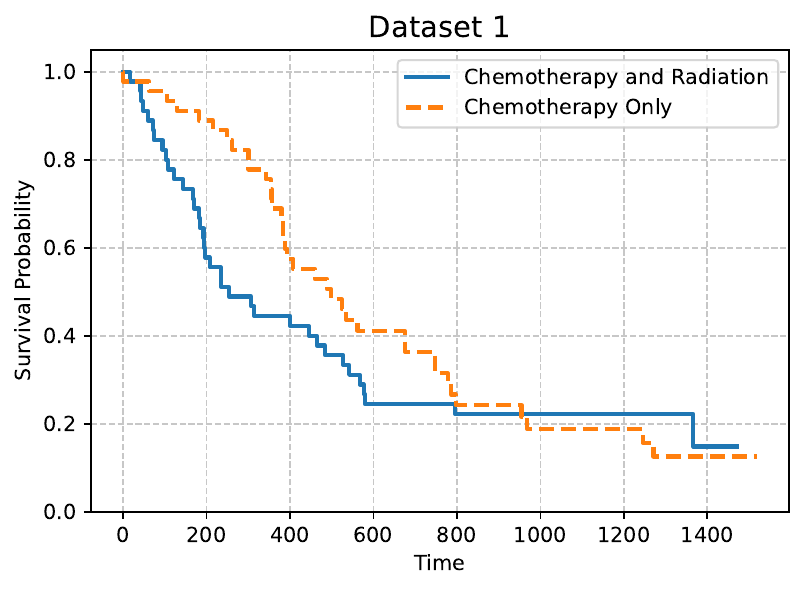}&
\includegraphics[width=8cm, height=6.5cm]{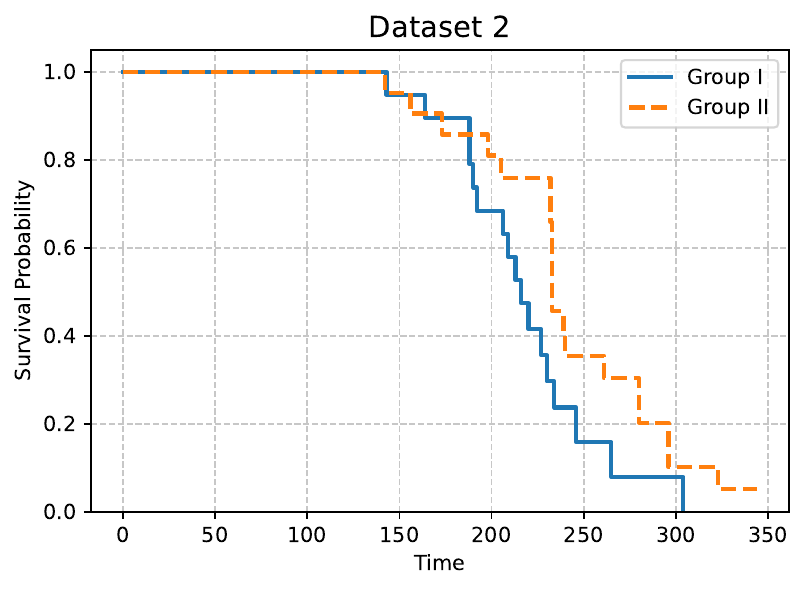}\\
\includegraphics[width=8cm, height=6.5cm]{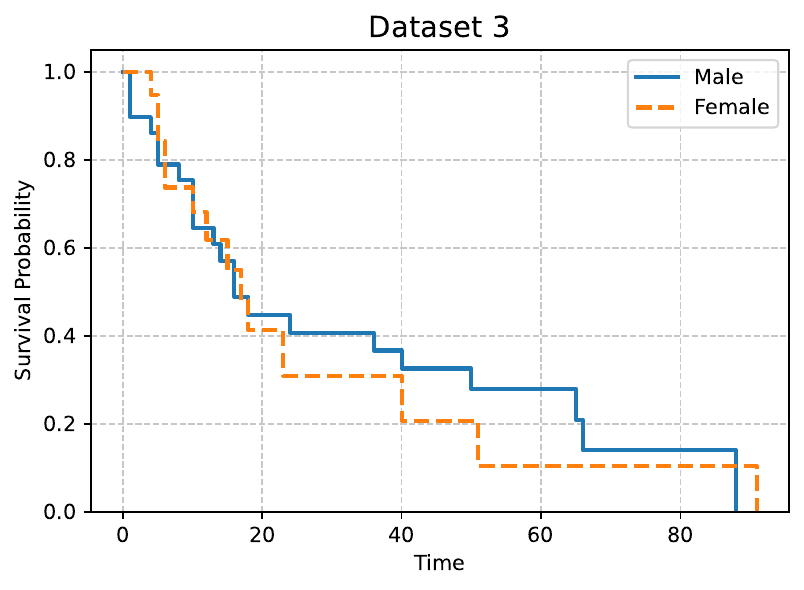}&
\includegraphics[width=8cm, height=6.5cm]{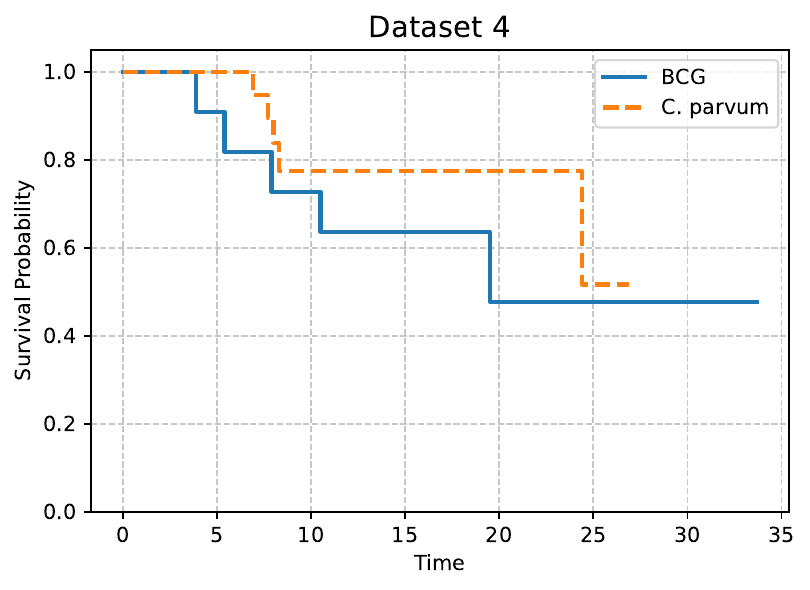}\\
\end{tabular}
\caption{Survival distributions for real-life datasets.} 
\label{fig1-survival}%
\end{figure}

To evaluate the practical performance of the proposed test, we applied it to four real-world datasets from published studies and compared the results with those from established nonparametric tests: the Gehan, Cox-Mantel, Log-rank, and Peto-Peto's tests. The survival data for these examples are provided in Table~\ref{tab9-realapp}, and the corresponding $p$-values for all tests are summarized in Table~\ref{tab10-dataset-p-value}. Graphical representations of the survival curves for these datasets are shown in Figure~\ref{fig1-survival}.
\\\\
\textbf{Dataset 1 (Gastric carcinoma trial):} The first dataset is from a clinical trial by \cite{Stablein1981} investigating treatments for locally advanced nonresectable gastric carcinoma. The study comprised 90 patients randomized to either a combination of chemotherapy and radiation ($n_1=45$) or chemotherapy alone ($n_2=45$). The $p$-values in Table~\ref{tab10-dataset-p-value} reveal a clear divergence in test conclusions. The Gehan ($p = 0.0294$), Peto-Peto ($p = 0.0334$), and the proposed test ($p = 0.0014$) all indicate a statistically significant difference between the two treatment survival distributions. In contrast, the Cox-Mantel ($p = 0.2998$) and Log-rank ($p = 0.3018$) tests fail to reject the null hypothesis at conventional significance levels.

\begin{table}[!t]
\centering
\setlength{\tabcolsep}{3pt} 
\caption{Real-life datasets.}
\begin{tabular}{lccccccccc}
\hline
\multicolumn{10}{@{}l}{Dataset 1: The survival time of clinical trials for chemotherapy and a combination of chemotherapy and}\\
\multicolumn{10}{@{}l}{~~~~~~~~~~~~~~~~radiation therapy.}\\\hline
Chemotherapy and Radiation: &17&42&44&48&60&72&74&95&103\\
&108&122&144&167&170&183&185&193&195\\
&197&208&234&235&254&307&315&401&445\\
&464&484&528&542&567&577&580&795&1366\\
&855+&882+&892+&1031+&1033+&1306+&1335+&1452+&1472+\\
Chemotherapy: &1&63&105&129&182&216&250&262&301\\ 
&301&342&354&356&358&380&383&383&388\\ 
&394&408&460&489&499&524&535&562&675\\ 
&676&748&748&778&786&797&955&968&1245\\
&1271&381+&529+&945+&1180+&1277+&1397+&1512+&1519+\\\hline
\multicolumn{10}{@{}l}{Dataset 2: The survival time of rats exposed to carcinogen DMBA.}\\\hline
Group-I:&143&164&188&188&190&192&206&209&213\\
&216&220&227&230&234&246&265&304&216+\\
&244+&&&&&&&&\\
Group-II:&142&156&173&198&205&232&232&233&233\\
&233&233&239&240&261&280&280&296&296\\
&323&204+&344+&&&&&&\\\hline
\multicolumn{10}{@{}l}{Dataset 3: the survival time of patients (gender-wise) on multiple myeloma.}\\\hline
Male:&1&1&1&4&5&5&8&10&10\\
&10&13&14&16&16&18&24&36&40\\
&50&65&66&88&3+&10+&15+&40+&52+\\
&56+&76+&&&&&&&\\
Female:&4&5&5&6&6&10&12&15&17\\
&18&23&40&51&91&7+&11+&12+&18+\\
&18+&&&&&&&&\\ \hline
\multicolumn{10}{@{}l}{Dataset 4: The survival time for 30 resected melanoma patients.}\\\hline
BCG:&3.9&5.4&7.9&10.5&19.5&16.6+&16.9+&17.1+&23.8+\\
&33.7+&33.7+&&&&&&&\\
C. parvum:&6.9&7.7&8&8.3&24.4&7.8+&8.2+&8.2+&10.8+\\
&11+&12.2+&12.5+&14.8+&16+&18.1+&21.4+&23+&24.8+\\
&26.9+&&&&&&&&\\\hline
\end{tabular}
\label{tab9-realapp}%
\end{table}

This pattern of results is highly informative regarding the nature of the treatment effect. The significant findings from the Wilcoxon-type tests, which assign greater weight to early event times, suggest that the survival curves diverge early in the follow-up period. The insignificant result from the Log-rank test, which is optimal under the proportional hazards assumption, indicate that the treatment effect may not be constant over time. The strong signal from the proposed test ($p = 0.0014$) demonstrates its enhanced sensitivity to the specific alternative hypothesis present in these data, likely capturing an early separation in survival that the Log-rank test misses.\\\\
\textbf{Dataset 2 (Vaginal cancer in rats):} The second dataset, from \cite{Pike1966}, involves an experiment on vaginal cancer in rats exposed to the carcinogen DMBA, with two groups distinguished by their pretreatment regime. The $p$-values in Table~\ref{tab10-dataset-p-value} show that none of the tests reject the null hypothesis of identical survival functions at the \(\alpha = 0.05\) level. However, the Cox-Mantel ($p = 0.0755$) and the proposed test yield the lowest $p$-values ($p = 0.0789$), suggesting a marginal, albeit insignificant, difference between the groups. This contrasts with the other tests, which have $p$-values above 0.09. This indicates that the proposed test may be more sensitive to the subtle differences present in this dataset, even if they do not reach formal statistical significance.\\\\
\textbf{Dataset 3 (Multiple myeloma patients):} The third dataset, originally from \cite{Krall1975}, contains survival times for 48 multiple myeloma patients aged 50–80 years, with the aim of investigating the effect of gender on survival. As shown in Table~\ref{tab10-dataset-p-value}, all five tests yield large $p$-values ($p \ge 0.8498$), providing no evidence to suggest a difference in survival between male and female patients. This unanimous conclusion across all methods is consistent with the graphical representation of the survival curves in Figure~\ref{fig1-survival}, which shows considerable overlap throughout the study period.\\\\
\textbf{Dataset 4 (Melanoma immunotherapy trial):} The fourth dataset, from \cite{Lee2003}, compares two immunotherapies—BCG and C. parvum—for their ability to prolong remission in 30 resected melanoma patients. The $p$-values in Table~\ref{tab10-dataset-p-value} show that the Gehan, Cox-Mantel, Log-rank, and Peto-Peto tests all yield insignificant results $(p \ge 0.3183)$. In contrast, the proposed test yields a $p$-value of 0.0843, which, although not significant at the $0.05$ level, is substantially lower and suggests a potential difference between the treatments. This finding aligns with the visual assessment of the survival curves in Figure~\ref{fig1-survival}, which appear to separate, suggesting that the proposed test may be more adept at detecting the underlying survival difference in this small sample.

\begin{table}[!t]
\centering
\caption{$p$-values for the real-life data examples.}
\begin{tabular}{lcccccccccc}
\hline
\multicolumn{11}{@{}l}{~~~~~~~~~~~~~~~~~~~~~~~~~~~~~~~~~~~~~~~~~~~~~~~~~~~~~Test statistic}\\
\cline{3-11}
Dataset&&Gehan&&Cox-Mantel&&Log-rank&&Peto-Peto&&Proposed \\ \hline
Dataset 1&&0.0294&&0.2998&&0.3018&&0.0334&&0.0014\\ 
Dataset 2&&0.0975&&0.0755&&0.0917&&0.0920&&0.0789\\
Dataset 3&&0.9907&&0.8498&&0.8580&&0.9602&&0.9366\\
Dataset 4&&0.3183&&0.3873&&0.3887&&0.3352&&0.0843\\
\hline
\end{tabular}
\label{tab10-dataset-p-value}%
\end{table}  

\section{Conclusion}\label{concl}
This paper has introduced a new nonparametric test for comparing two survival distributions with right-censored data. The test addresses a critical weakness of existing methods—their dependence on the unknown alternative hypothesis—by leveraging a unique decomposition of the data into uncensored and censored observations. The resulting composite statistic automatically synthesizes evidence from both the event times and the censoring process, creating a robust omnibus procedure.

Simulation results confirm that the test provides a compelling solution to the practitioner's dilemma of test selection. It consistently controls the Type I error rate and delivers a power profile that is stable and high across a diverse spectrum of scenarios. Unlike traditional tests whose power can diminish severely under non-proportional hazards or high censoring, the proposed method remains effective, particularly excelling in the challenging context of crossing survival curves. Its performance aligns with its design goal: to offer a single, powerful test whose power approximates the best one could hope to achieve by correctly choosing between the log-rank and Wilcoxon tests a priori.

Given its distribution-free nature, computational simplicity, and robust power, the proposed test is a highly valuable addition to the survival analyst's toolkit. It is especially recommended for exploratory analyses, studies with unknown hazard patterns, or any application where robustness to complex survival differences and censoring is paramount.

\section*{Conflicts of interest}
No conflicts of interest is reported.

\section*{Data availability statement} 
The datasets are provided within the manuscript.

\section*{Funding statement} 
No grant received for this research.

\section*{Ethical statement}
The research work has not been previously published or submitted in any form.

\section*{Acknowledgments} 
We acknowledge the support of the large language models for language editing.

\section*{Code availability statement} 
The codes developed for this work can be accessed by \href{https://github.com/touqeerahmadunipd/Novel-NPtest}{https://github.com/touqeerahmadunipd/Novel-NPtest}

\end{document}